\documentclass[3p,times,procedia]{elsarticle}
\usepackage{nupha_ecrc}
\usepackage{mathtools}
\usepackage{commath}
\usepackage{graphicx}
\usepackage{lineno}

\volume{00}

\firstpage{1}

\journalname{Nuclear Physics A}

\runauth{}

\jid{nupha}

\jnltitlelogo{Nuclear Physics A}

\usepackage{amssymb}

\usepackage[figuresright]{rotating}

\begin{document}
\begin{frontmatter}



\dochead{XXVIth International Conference on Ultrarelativistic Nucleus-Nucleus Collisions\\ (Quark Matter 2017)}


\title{Charge asymmetry dependence of anisotropic flow in pPb and PbPb collisions with CMS experiment}


\author{Sang Eon Park on behalf of the CMS Collaboration}

\address{Physics \& Astronomy MS-61, Rice University, Houston, US, 77005}
\ead{sp35@rice.edu}
\begin{abstract}
In nucleus-nucleus collisions, the linear dependence found for the elliptic flow harmonic of both positive or negative charged particles as a function of event charge asymmetry is predicted by the phenomenon known as the Chiral Magnetic Wave (CMW) due to its induced electric quadrupole moment. Here, the event charge asymmetry $A_{\rm ch}$ is defined as $\frac{N_{+}-N_{-}}{N_{+}+N_{-}}$, where $N_{+}$ and $N_{-}$ are the number of positive and negative charged particles, respectively. However, other scenarios are also possible and may provide alternative explanations for the experimental results. New measurements of elliptic ($v_{\rm 2}$) and triangular ($v_{\rm 3}$) flow for positive and negative charged particles as a function of $A_{\rm ch}$ in pPb and PbPb collisions at $\sqrt{s_{_{NN}}} = 5.02~\mathrm{TeV}$ are presented, using data collected by the CMS experiment during the LHC runs 1 and 2. The slopes and intercepts of the charged-dependent $v_{n}$ harmonics vs. $A_{\rm ch}$ are directly compared for pPb and PbPb collisions with similar charged-particle multiplicities, where a strong CMW effect is not expected in very high multiplicity pPb events. Moreover, a comparison is made of the slope parameters between $v_{2}$ and $v_{3}$ harmonics normalized by the inclusive charge particle $v_{n}$ in PbPb collisions as a function of centrality. These results provide a means to discriminate between the CMW and other scenarios such as local charge conservation as possible explanations for the observed charge dependent behavior.
\end{abstract}

\begin{keyword}
CMW \sep Chiral Magnetic Wave \sep LCC \sep local charge conservation \sep small systems \sep chiral magnetic effect

\end{keyword}

\end{frontmatter}


\section{Introduction}
\label{introduction}

In relativistic nucleus-nucleus collisions, the nonzero axial chemical potential, induced by the chiral anomaly with an imbalance of left- and right-handed quarks, can produce an electric current along the magnetic field that is produced by the spectator protons~\cite{Kharzeev:2004ey,Kharzeev:2007jp,Kharzeev:2015znc}. The phenomenon of electric charge separation along the axis of magnetic field is known as the chiral magnetic effect (CME)~\cite{Kharzeev:2004ey}. The chiral separation effect (CSE) is a similar process, where the chiralities can be separated along the magnetic field direction due to the presence of the electric charges. The CME and CSE together form a collective excitation, known as the chiral magnetic wave (CMW). The propagation of the CMW produces an electric quadruple moment in the system, where additional positive (negative) charges are accumulated away from (close to) the reaction plane~\cite{Burnier:2011bf}. 

It has been proposed that this electric quadruple moment in the quark-gluon plasma fireball can induce an effect on the second-order azimuthal anisotropic coefficient (known as ``elliptic flow'', $v_2$) of the final-state particles, which then exhibits a negative (positive) correlation to
event charge asymmetry, $A_{\rm ch} \equiv \frac{N_{+}-N_{-}}{N_{+}+N_{-}}$, for positive- (negative-) charged particles~\cite{Burnier:2011bf}.
Here, $N_{+}$ and $N_{-}$ denote the number of positive- and negative-charged particles in each event. Therefore, the elliptic flow coefficients ($v_{\rm 2}$) of positive- and negative-charged particles show a linear dependence on $A_{\rm ch}$ as follows, $v_{2,\pm} = v_{2,\pm}^{base} \mp rA_{\rm ch}$ ,where $v_{2,\pm}^{base}$ represents the $v_{\rm 2,\pm}$ value when $A_{\rm ch} = 0$ and the $r$ denotes the slope parameter. 

Recent measurements of charge asymmetry dependence of $v_{2,\pm}$ in AA collisions made at the BNL's RHIC and the CERN's LHC provide evidence in line with the CMW mechanism~\cite{Abelev:2012pa, Adamczyk:2015eqo, Adam:2015vje}. However, the interpretation of the results still remains highly debated. For example, the local charge conservation effect has been shown to qualitatively explain the charge-dependent $v_2$ data as a function of event charge asymmetry~\cite{Bzdak:2013yla}, where the signal arises
from the interplay of finite detector acceptance and the dependence of $v_{n}$ on pseudorapidity ($\eta$) and/or transverse momentum ($p_{\mathrm{T}}$) 
for charged particles from decays of resonances, resulting in a correlation between $p_{\mathrm{T}}$ ($\eta$)-averaged $v_n$ and $A_{\rm ch}$.

This contribution presents the measurements of charge asymmetry dependence of $v_{\rm 2}$ coefficients in pPb collisions, and both $v_{\rm 2}$ and $v_{\rm 3}$ coefficients in PbPb collisions at $\sqrt{s_{_{NN}}} = 5.02~\mathrm{TeV}$, using data collected with the CMS detector at the LHC~\cite{Chatrchyan:2008aa}. In recent years, CMS has reported the flow-like behavior and its collective nature in high-multiplicity pp and pPb collisions, in terms of radial flow and final-state azimuthal anisotropy correlations, which are similar to that in AA collisions~\cite{Khachatryan:2016yru,Khachatryan:2010gv,CMS:2012qk,Khachatryan:2014jra,Chatrchyan:2013nka,Khachatryan:2015waa,Dusling:2015gta,Khachatryan:2016txc,Khachatryan:2015lva}. However, the CMW contribution to any charge asymmetry dependent $v_n$ signal is expected to be very small in a pPb collision compared to a PbPb collision with similar event multiplicity. This is mainly due to the fact that the induced magnetic field in pPb is much reduced in strength, and most importantly, oriented randomly with respect to the event plane~\cite{Khachatryan:2016got}. Furthermore, a direct comparison of $A_{\rm ch}$-dependent $v_2$ and $v_3$ coefficients in PbPb collisions can provide important evidence to differentiate between the CMW and local charge conservation (LCC) mechanisms.

\section{Results}
\label{results}

\begin{figure*}[thb]
\centering
\mbox{\includegraphics[width=2.4in]{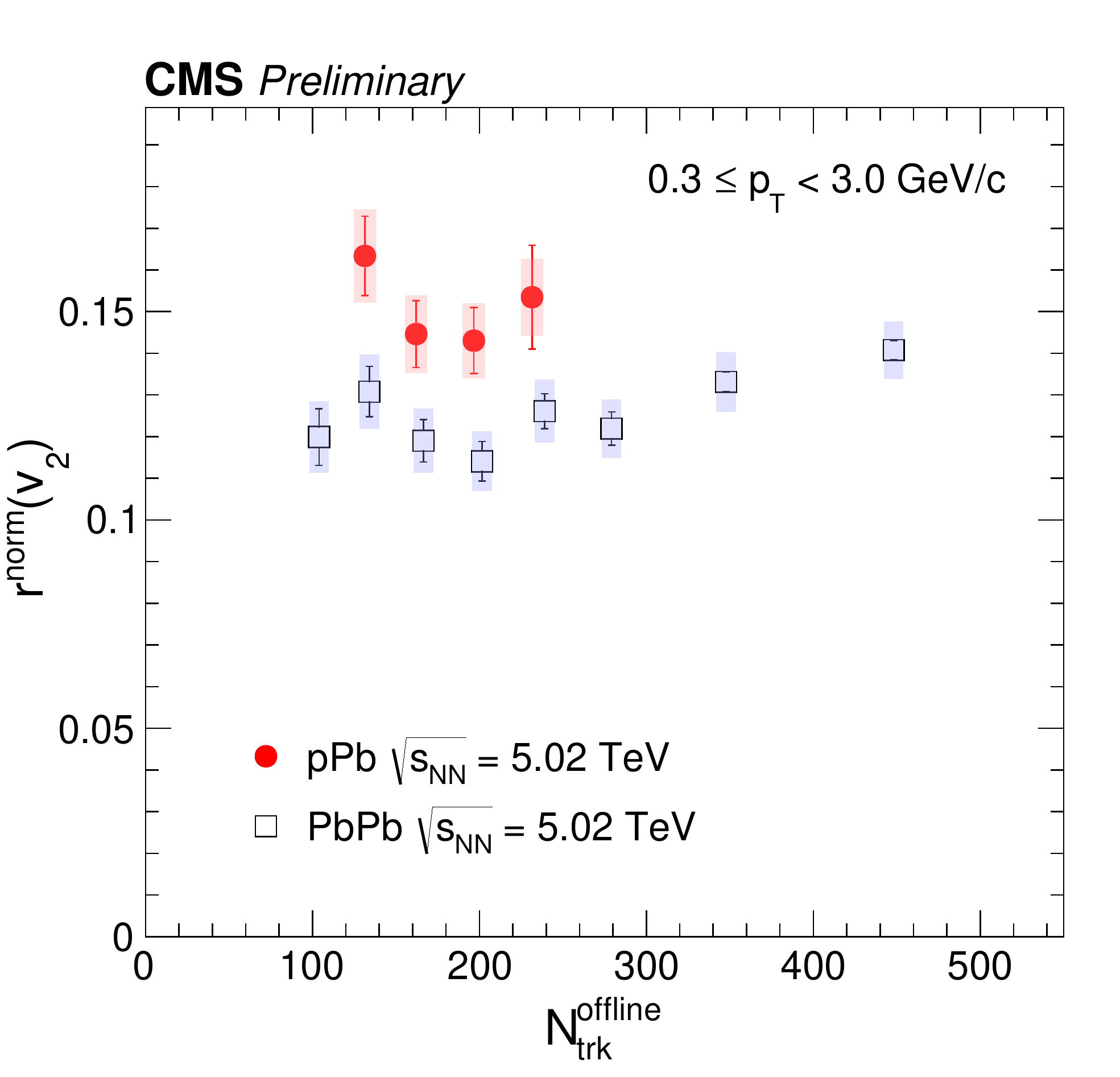}}
\mbox{\includegraphics[width=2.4in]{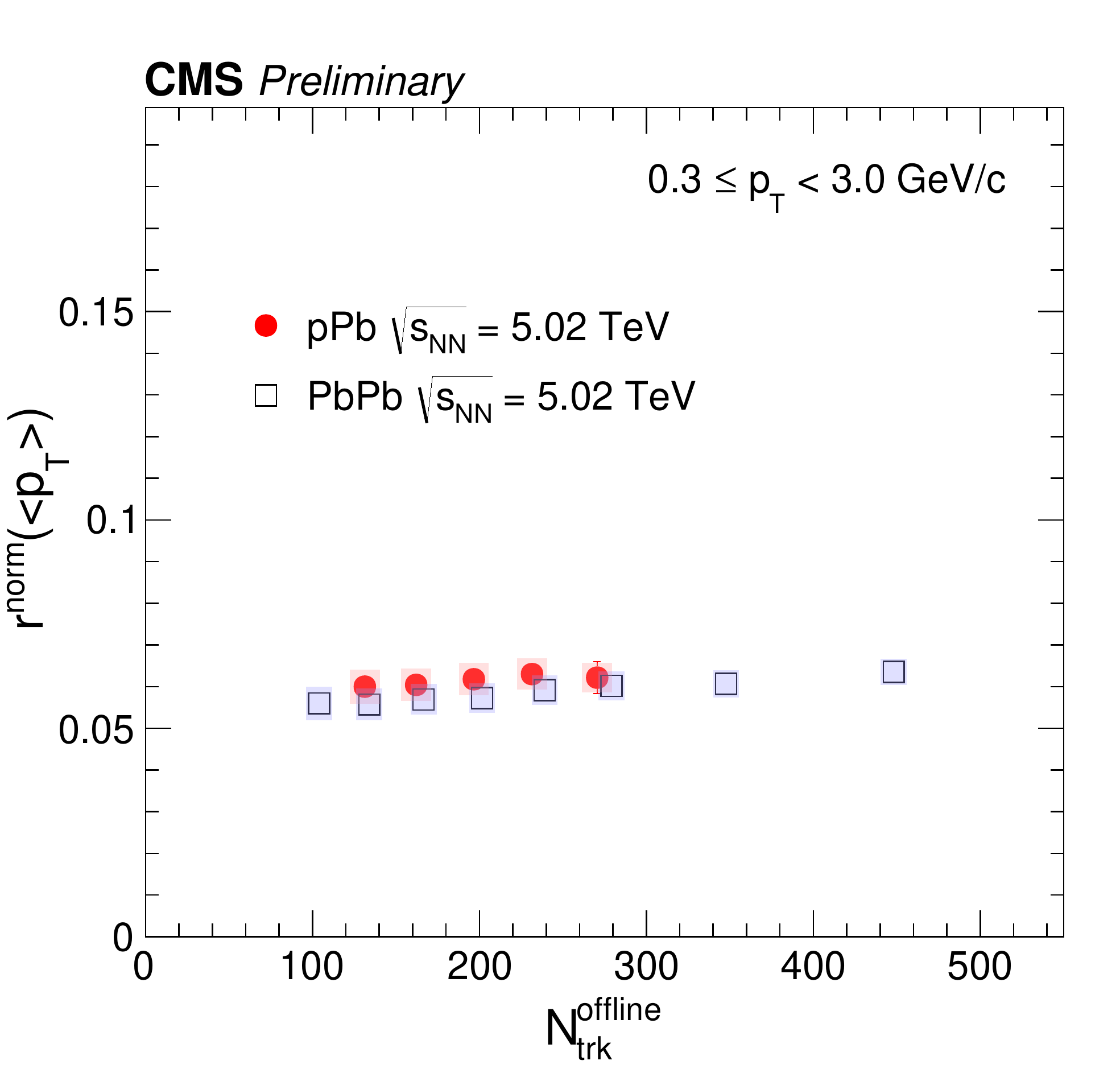}}
\caption{ 
The linear slope for $v_2$ (left) and $\left< p_{\mathrm{T}} \right>$ (right)
parameters $r^{\rm norm}(v_2)$ and $r^{\rm norm}({\left< p_{\mathrm{T}} \right>})$ as a function of event multiplicity in pPb and PbPb 
collisions at $\sqrt{s_{_{NN}}}$ = 5.02 TeV. Statistical and systematic uncertainties
are indicated by the error bars and shaded regions, respectively~\cite{CMS-PAS-HIN-16-017}}
\label{fig:figure3}
\end{figure*}

Figure~\ref{fig:figure3} (left) shows the extracted slope parameters for $v_2$, averaged over $0.3<p_{\mathrm{T}}<3$GeV/c, as a function of event multiplicity in pPb and PbPb collisions at $\sqrt{s_{_{NN}}}$ = 5.02TeV. The linear slope parameter, $r^{\rm norm}(v_2)$, is extracted by linearly fitting the normalized $v_{\rm 2}$ difference, $(v^{-}_{2}-v^{+}_{2})/(v^{-}_{2}+v^{+}_{2})$, as a function of the corrected $A_{\rm ch}$. 

In the overlapping multiplicity range between pPb and PbPb, slope parameter for pPb was slightly larger than that of PbPb. As discussed previously, no CMW effect is expected in high-multiplicity pPb events. Therefore, significant slope parameter observed in pPb compared to that of PbPb poses a considerable challenge to the mechanism of the CMW.

The event-averaged $p_{\mathrm{T}}$ values ($\left< p_{\mathrm{T}} \right>$) for positive and negative charged particles as a function of $A_{\rm ch}$ show the same pattern as in $v_{\rm 2}$ case, therefore we can extract the slope parameter $r^{\rm norm}({\left< p_{\mathrm{T}} \right>})$ for the event averaged $p_{\mathrm{T}}$ values the same way, by linearly fitting the normalized $\left< p_{\mathrm{T}} \right>$  difference as a function of the corrected $A_{\rm ch}$. Fig.~\ref{fig:figure3} (right) shows the $r^{\rm norm}({\left< p_{\mathrm{T}} \right>})$ for pPb and PbPb
collisions at $\sqrt{s_{_{NN}}}$ = 5.02 TeV as a function of $N^{\rm offline}_{\rm trk}$. Non-zero slope parameters for $\left< p_{\mathrm{T}} \right>$ are observed for all measured multiplicities, which account for about 50\% of the observed slope for $v_2$ at $0.3<p_{\mathrm{T}}<3$GeV/c. 
This observation is in line with the predicted local charge conservation effect.

\begin{figure*}[thb]
\centering
\mbox{\includegraphics[width=5in]{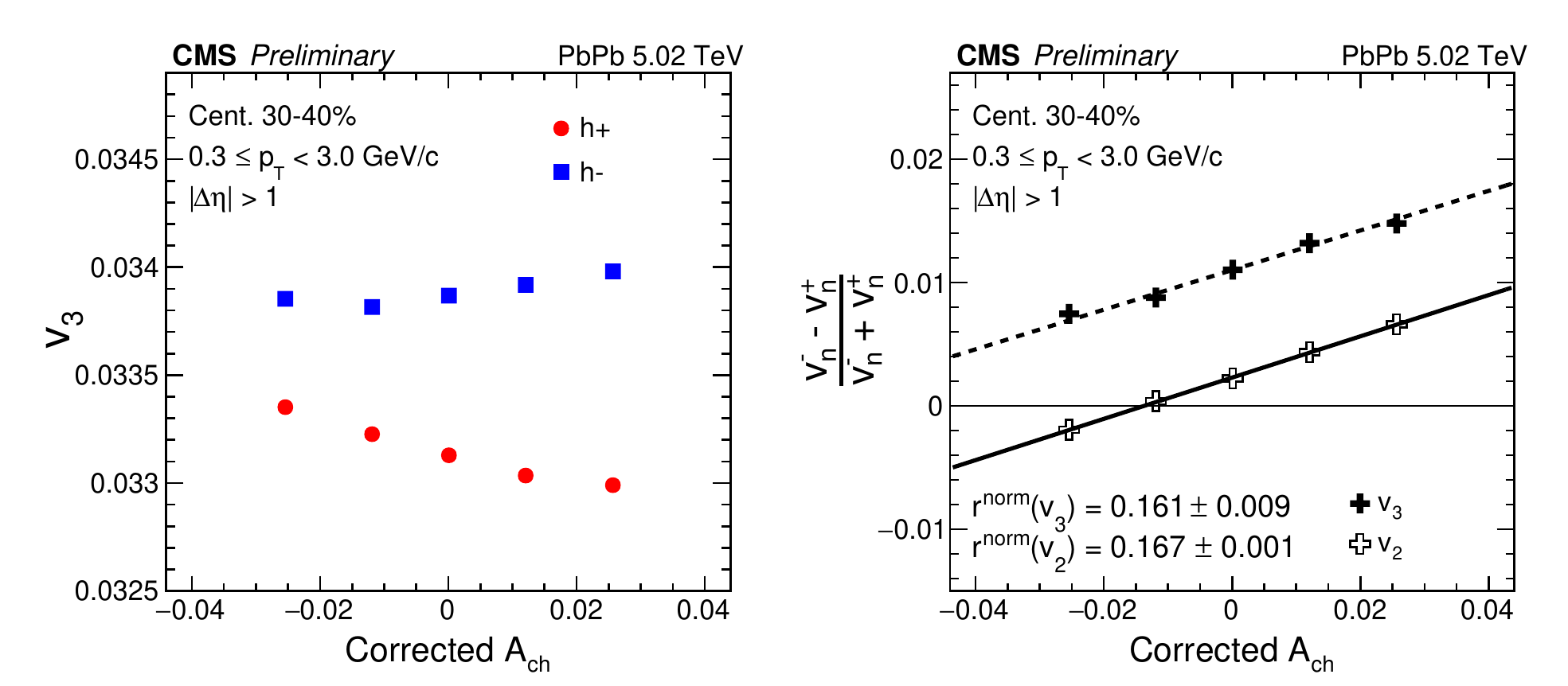}}
\caption{
The triangular flow $v_3$ for positive- and negative-charged particles (left) 
and the normalized difference in $v_{n}$, $(v^{-}_{n} - v^{+}_{n})/(v^{-}_{n} + v^{+}_{n})$, 
for $n=2$ and $n=3$ (right) as a function of corrected event charge asymmetry 
for 30--40\% centrality of PbPb collisions at $\sqrt{s_{_{NN}}}$ = 5.02 TeV.
Statistical uncertainties are covered by the data points~\cite{CMS-PAS-HIN-16-017}.
}
\label{fig:figure4}
\end{figure*}

The charge asymmetry dependence of triangular flow $v_{3}$ for positive-
and negative-charged particles is also studied in PbPb
collisions at $\sqrt{s_{_{NN}}}$ = 5.02 TeV, shown in Fig.~\ref{fig:figure4} (left)
for 30--40\% centrality range. Similar to the case of $v_2$, the $v^{+}_{3}$ ($v^{-}_{3}$) 
decreases (increases) as $A_{\rm ch}$ increases.

\begin{figure*}[thb]
\centering
\mbox{\includegraphics[width=2.7in]{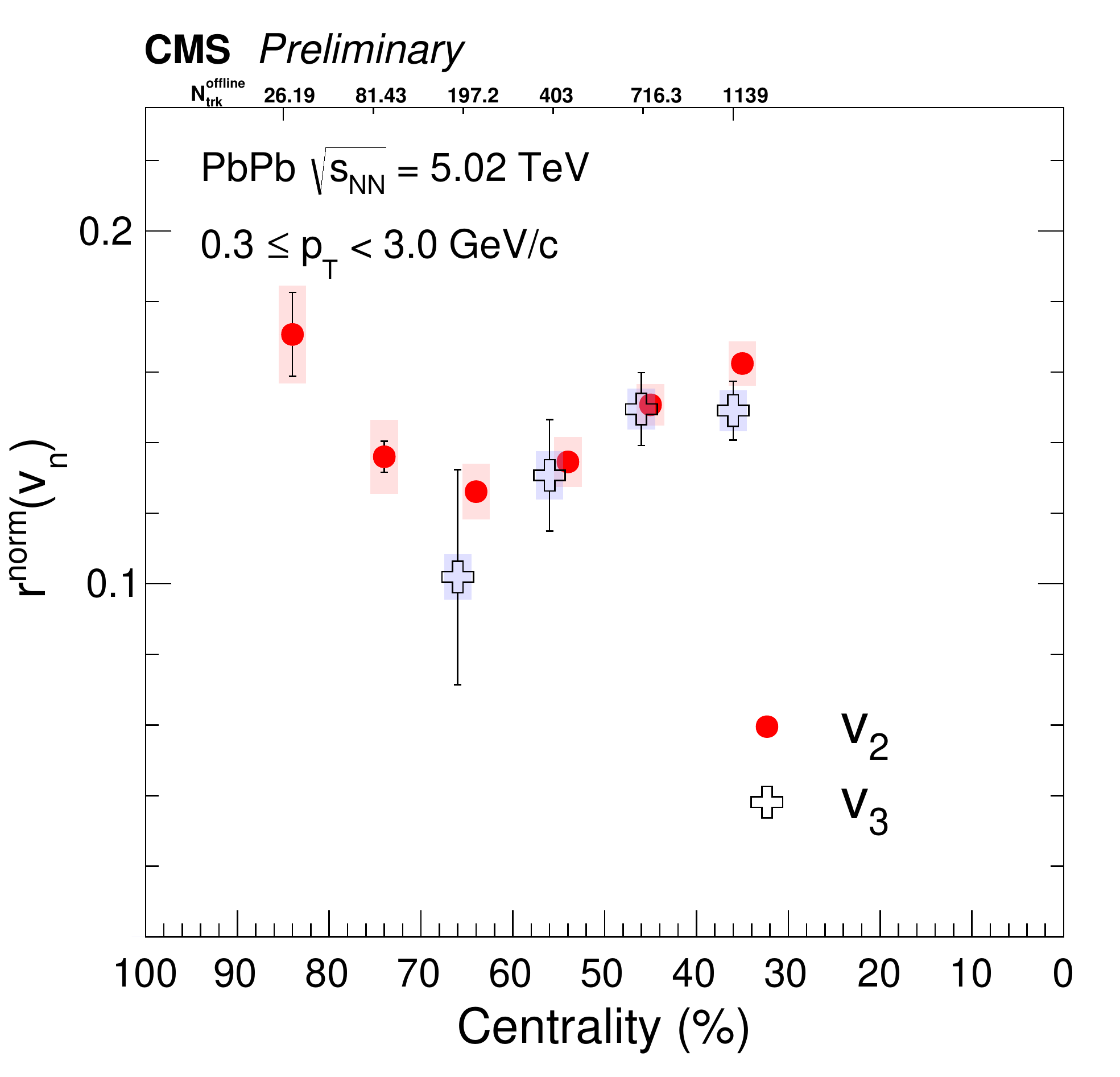}}
\caption{
The linear slope ($r^{\rm norm}(v_n)$) 
parameter for elliptic ($n=2$) and triangular ($n=3$) flow as a function of centrality 
for PbPb collisions at $\sqrt{s_{_{NN}}}$ = 5.02 TeV. Statistical and systematic uncertainties
are indicated by the error bars and shaded regions, respectively~\cite{CMS-PAS-HIN-16-017}.
}
\label{fig:figure5}
\end{figure*}

The normalized $v_3$ difference, 
$(v^{-}_{3}-v^{+}_{3})/(v^{-}_{3}+v^{+}_{3})$, is then plotted as a function of 
the corrected $A_{\rm ch}$ values in Fig.~\ref{fig:figure4} (right), and also
compared to that for $v_2$. As shown, the slopes $r^{\rm norm}(v_2)$
and $r^{\rm norm}(v_3)$ agree well within statistical uncertainties. The $r^{\rm norm}(v_n)$ for both $n=2$ and $n=3$
as a function of centrality of PbPb collisions at $\sqrt{s_{_{NN}}}$ = 5.02 TeV are shown 
in Fig.~\ref{fig:figure5} for the centrality range of 30--80\%.
Over the centrality range studied, the slope parameters for elliptic and triangular flow are consistent with each other within uncertainties.

In the CME and CSE, the charges are expected to separate with respect to the reaction 
plane ($\Psi_{RP}$), or the second-order event plane ($\Psi_{2}$). Therefore CMW effect is highly suppressed with respect to the third-order 
event plane ($\Psi_{3}$), leading to a vanishing slope parameter $r^{\rm norm}(v_3)$. The similar $r^{\rm norm}(v_2)$ and $r^{\rm norm}(v_3)$ values between elliptic and triangular flow observed in the data again indicate an underlying physics mechanism not related to the CMW. This behavior is qualitatively consistent with the prediction from the local charge conservation effects~\cite{Bzdak:2013yla}. 

\section{Summary}
\label{summary}

In summary, the charge-dependent elliptic ($v_{\rm 2}$) and triangular ($v_{\rm 3}$) flow coefficients in pPb and PbPb have been measured as a function of event charge asymmetry at $\sqrt{s_{_{NN}}} = 5.02~\mathrm{TeV}$. The normalized slope parameters of $v_{\rm 2}$ in pPb and PbPb collisions are found to be similar at the same event charged-particle multiplicities. Similar charged asymmetry dependence is also observed for the event-averaged $p_{\mathrm{T}}$ values of positive- and negative-charged particles. Moreover, in PbPb collisions, the normalized slope parameters of the $v_{\rm 2}$ and $v_{\rm 3}$ coefficients show similar magnitudes in various centralities. All these observations would not be expected from the mechanism of chiral magnetic wave, but are qualitatively consistent with predictions from the local charge conservation mechanism.





\bibliographystyle{elsarticle-num}
\bibliography{HIN16009}







\end{document}